# Influence of Permeability Anisotropy and Layered Heterogeneity on Geothermal Energy Battery Storage

Palash Panja[1,2], John McLennan[1], Sidney Green[3]

[1]Department of Chemical Engineering, University of Utah, Salt Lake City, Utah, the United States

[2]Energy & Geoscience Institute, University of Utah, Salt Lake City, Utah, the United States

[3]Enhanced Production Inc. & Research Professor, University of Utah, Salt Lake City, Utah, the United States

## Abstract

The Geothermal Battery Energy Storage concept has been proposed to provide large-scale heat storage when solar radiance is available, to be later recovered for economic benefit. The concept uses solar radiance to heat water on the surface which is then injected into a suitable subsurface formation. This hot water elevates the ambient formation temperature creating a high-temperature geothermal reservoir acceptable for geothermal electricity generation or direct heat applications. The process uses produced/injected, connate formation water and thus neither freshwater nor surface storage or disposal of water is required. This concept has been previously presented in several publications and presentations.

Calculations of reservoir temperature and pressure profiles for injection and production in isotropic and homogeneous reservoirs have been published previously by the authors. These calculations have shown that a small volume of rock mass is required for the heat storage reservoir, of the order of tens of meters radius from an injection well in a reservoir of one-hundred meters thickness. With this small rock mass volume, locations away from fractures, faults, and inclusions are possible. It was shown that over ninety percent of heat can be recovered for certain reservoirs.

The previous calculations for the Geothermal Battery Energy Storage considered only isotropic and homogeneous reservoir formation properties. However, even in a small rock mass volume, considering sedimentary depositional environments and superimposed tectonics, the rock permeability may be anisotropic and heterogeneous with reservoir layers of different properties. Calculations are presented here considering anisotropic permeabilities, and layered heterogeneous permeabilities i.e., formations with horizontal layers of different permeabilities. Such reservoir properties create non-symmetrical temperature and pressure profiles away from a well, which is critical for well layout and planning for injection and production.

## Introduction

Industrial-scale storage of heated water by solar irradiance in a geothermal system is a challenge in the emergence of new technologies to harness renewable energy. Hybridizing a power plant with solar power and thermal storage is a potential solution for large-scale energy storage in the Earth during periods of adequate solar radiance [1-3]. The NSF funded SedHeat Geothermal Energy Project [4, 5] recommended injecting hot water derived from solar thermal energy into low-quality geothermal sedimentary basins. Recently, a team led by Idaho National Laboratory[6] showed that high permeability and porosity formations could potentially provide storage for solar heated water, with subsequent cyclic production and reinjection. Only a few studies consider anisotropy in a geothermal reservoir. Researchers from Sandia National Laboratories [7] studied the effects of vertical anisotropy in fracture permeability using fractured continuum model (FCM) in an enhanced geothermal system (EGS) in granite formation. They showed an insignificant difference in heat extraction with small fracture spacing compared to isotropic permeability. Although, more heat was extracted with large fracture spacing and vertical anisotropy in permeability. Others [8, 9] also studied the effects of layered heterogeneity in Enhanced Geothermal System (EGS). In our previous study [10], heat storage performance in a homogeneous (and isotropic in horizontal permeability) reservoir was evaluated. We showed that geothermal battery storage is a potential technology in storing hot water in high permeability and porosity homogeneous formations during periods of adequate solar radiance. The water can be produced for power generation when necessary and subsequently reinjecting the same water after heating. Many factors that control the recovery of heat were discussed in the same study. The objective of this study is to investigate the influences of anisotropy, and layered heterogeneities in permeability on heat recovery, temperature, and pressure profiles in geothermal porous media using huff-and-puff operation.

**Reservoir Model**

The heat and fluid flow in a porous media are simulated using a thermal simulator, STAR, from the Computer modeling group (CMG), Calgary, Canada. A three-dimensional model of the reservoir is created using a cartesian grid system. The insulating, relatively impermeable overburden at the top, the storage domain in the middle (formation), and the insulating underburden section at the bottom are the three segments of the model. The overburden and underburden are each 70 meters thick with 7 gridded layers in each section. The formation (middle segment) has 11 horizontal layers with a total thickness of 110 meters. Each layer has the same grid refinement in the x and y directions. 215 grids are assigned in the x-direction and y-direction for a total length of 200 meters in each direction. The grid sizes in the x and y directions vary with refined grids near the wellbore for numerical stability and accuracy. The total number of grid blocks in the model is 1155625.

A constant pressure boundary condition is employed. An analogy is to view the reservoir as being surrounded by an aquifer with pressure support. These constant pressure boundaries are maintained at 12 MPa which is the initial reservoir pressure. The overburden and underburden have different geologic and thermophysical properties. Geologic properties such as the porosity and permeability of the overburden and underburden sectors are low (2.5% and 100 nD, respectively) compared to the formation segment. Except for the anisotropy studies, the permeability in the x-direction ($k_x$) is the same as the permeability in the y-direction ($k_y$) and 10 times the permeability in the z-direction ($k_z$). The base parameters for overburden, formation, and underburden are listed in **Table 1** [10].

**Table 1:** List of base parameters for the formation, overburden and underburden[10]

| Parameters | Formation | Overburden/ Underburdon |
|---|---|---|
| Specific heat of rock (J/(Kg-K) | 930 | 770 |
| Thermal Conductivity (W/(m-K) | 2.5 | 1.05 |
| Density (kg/m$^3$) | 2000 | 2500 |
| Mean horizontal permeability, $k_x$ and $k_y$ (mD) | 100 | 0.0001 |
| Porosity (%) | 15 | 2.5 |
| Initial Temperature ($^O$C) | 120 | 120 |
| Initial Pressure (MPa) | 12 | 12 |
| Formation thickness (m) | 110 | 70 |

A suite of simulations was conducted to study the effects of horizontal and vertical permeability anisotropies and heterogeneities. Three horizontal anisotropies ($k_x/k_y$ =1, 2 and, 5), three vertical anisotropies ($k_x/k_z$ = 1, 2, 10), and four-layered heterogeneities are studied. The results are also compared with a homogeneous and horizontal isotropic permeability model. In the case of layered heterogeneity, each horizontal layer of the formation segment has different permeability as described later.

Hot water at 250°C from the surface facility is injected for 8 hours at a rate of 40 kg/s. The same well is then used to produce water for 10 hrs at a rate of 32 kg/s to assure the same amount of water production as injected. After the injection and production cycles of a total of 18 hours, well is shut-in for the rest of the day ( 6 hours) for thermal and pressure equilibrium inside the reservoir before the next cycle starts. The schedule is kept unchanged for each day. The calculations are continued for 30 cycles i.e., 30 days.

**Results**

In the following sections, results from the horizontal anisotropy, and layered are discussed separately. The results from vertical anisotropy are presented in the appendix. Emphasis is given more on the temperature and pressure distributions inside the formation after the end of injection and, production of the 30$^{th}$ cycle. The heat recoveries with the number of cycles (or days) are also calculated. Additionally, the bottom hole temperature and pressure versus the number of cycles are discussed in some cases.

**Effect of Horizontal Anisotropy**

The effects of horizontal permeability anisotropy, represented by the ratio of absolute permeability in the x-direction and the y-direction ($k_x/k_y$), are discussed. Because all 11 layers of formation are perforated, preferential horizontal flow can be assumed in x- and y-directions only. The permeability in the x-direction is kept at 100 mD and the permeability in the y-direction is varied as 100 mD ($k_x/k_y = 1$), 50 mD ($k_x/k_y = 2$) and 20 mD ($k_x/k_y = 5$). The pressure distributions at mid-height in the x-y plane at the end of injection in the 30$^{th}$ cycle are shown in **Figure 1** for three different horizontal anisotropies.

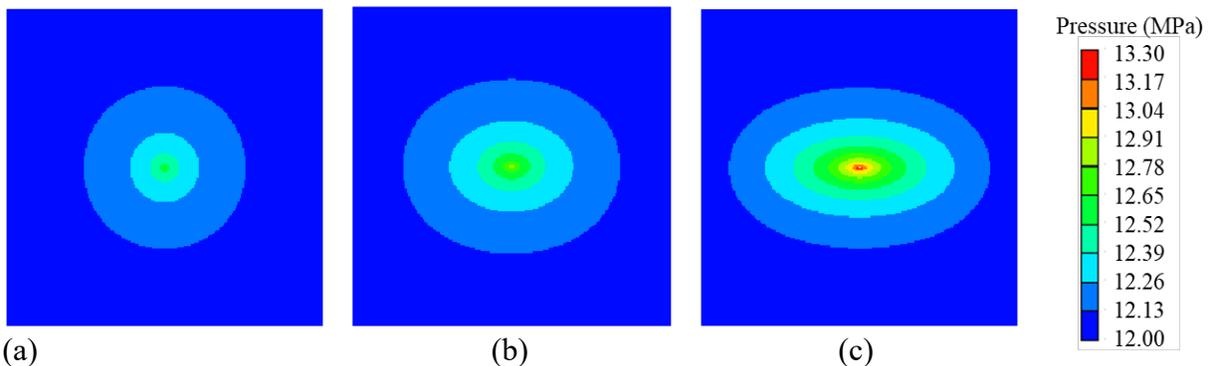

(a) (b) (c)
**Figure 1**: Spatial distribution of pressure (kPa) at the end of injection of the 30$^{th}$ cycle of operation on the x-y plane (200 meters x 200 meters) at the mid-height of the formation for horizontal anisotropy, $k_x/k_y$ of (a) 1 (b) 2 (c) 5

Recognizing that the same mass of water is injected in each case, the flow experiences more resistance in the higher horizontal anisotropy because equivalent permeability on the horizontal plane is reduced. These sorts of analyses are very common, where an analytic permeability equivalent is often specified as the square root of the sum of the squares of the orthogonal permeabilities. As anticipated, the pressure front moves farther in the direction of the higher absolute permeability. The pressure front travels farther in the x-direction as horizontal anisotropy increases. Therefore, the shape of the isobaric contours becomes more elliptic. Moreover, absolute pressure at the wellbore during injection also increases in the spatial distribution with the horizontal anisotropy. The temperature distribution after the end of injection in the 30th cycle is shown in **Figure 2**.

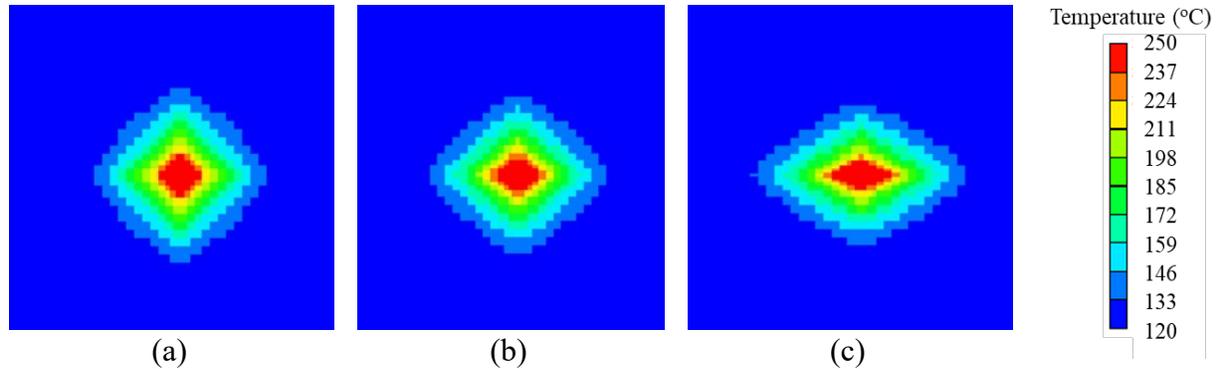

(a) (b) (c)

**Figure 2**: Spatial distribution of temperature (°C) at the end of injection during the 30$^{th}$ cycle of operation on the x-y plane (40 meters x 40 meters) at the mid-height of the permeable formation for horizontal anisotropy, $k_x/k_y$ of (a)1 (b) 2 (c) 5.

As explained above, the flow is preferential in the x-direction with increasing permeability anisotropy; the temperature has similar behavior. Since the temperature front is equivalent to (but lags) the pressure front, hot water is carried preferentially in the x-direction. The thermal front moves only a few meters (~25 meters only) compared to the pressure front (100 meters). Again, this is supported by legacy simulations, for example, Perkins and Gonzalez [11]. The thermal front lags behind the pressure front because the hot injected water (250 °C) quickly mixes with relatively cold reservoir water (120 °C) and the surrounding reservoir water is pushed towards the boundary. Along with the spatial distribution, pressure and temperature are also plotted along the x- and y-axes, as shown in **Figure 3**.

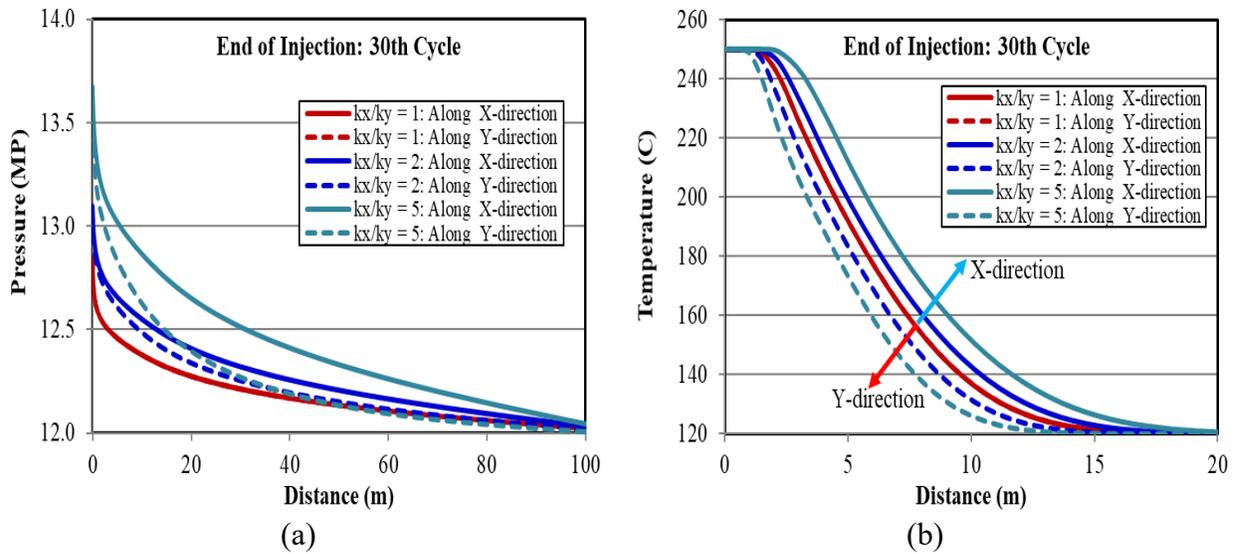

**Figure 3**: (a) Pressure and (b) temperature profiles at the end of injection of the 30$^{th}$ cycle of operation along the two perpendicular x and y coordinate axes at the mid-height of the formation for different horizontal anisotropies.

The temperature and pressure are the same in both directions (x and y) for the isotropic case; i.e., $k_x/k_y = 1$. This is due to the equal movement of the pressure or temperature front in the x and y directions for the same permeability. For higher anisotropy, the pressure and temperature fronts are always "ahead" in the x-direction compared to the y-direction. For example, for a horizontal anisotropy of $k_x/k_y = 5$, a 12.5 MPa pressure front reaches 31.3 meters in the x-direction and only 14.7 meters in the y-direction. Similarly, a 160$^O$C temperature front moves 9.1 meters in the x-direction and only 5.9 meters in the y-direction. The pressure distributions at mid-height in the x-y plane after the end of production in the 30th cycle are shown in **Figure 4** for three different horizontal anisotropies.

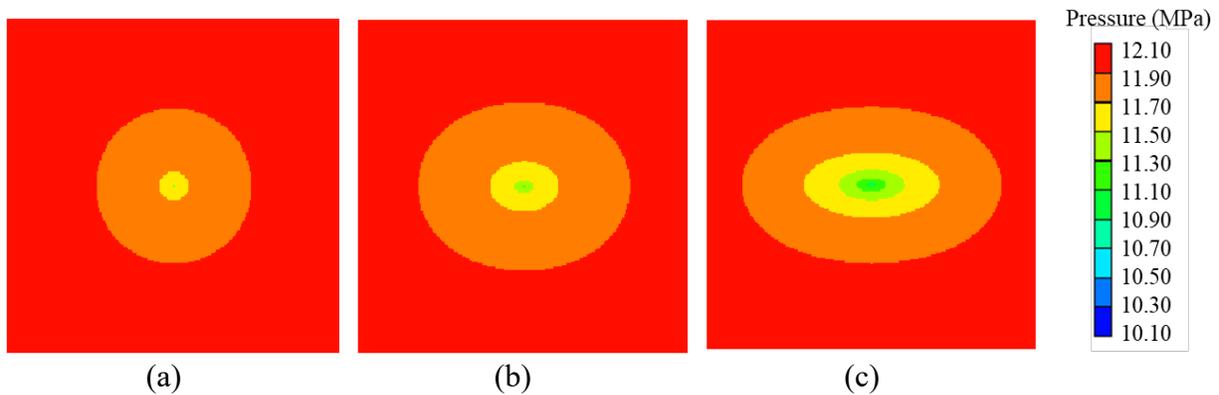

**Figure 4**: Spatial distribution of pressure (kPa) at the end of production of the 30$^{th}$ cycle of operation on the x-y plane (200 meters x 200 meters) at mid-height of the formation for horizontal anisotropy, $k_x/k_y$ of (a) 1 (b) 2 (c) 5.

The shapes of the pressure and temperature distributions after the end of production in the 30th cycle are similar. The pressure near-the wellbore is low and it increases towards the boundary of the reservoir. The isobaric contour becomes more elliptic with increasing horizontal anisotropy. The shape is a circle for the isotropic case ($k_x/k_y = 1$). The temperature front moves only a few meters. The shape is also elongated in the x-direction due to the higher permeability in this direction. The pressure and temperature along the x- and y-axes at the end of production in the 30th cycle are shown in **Figure 5**.

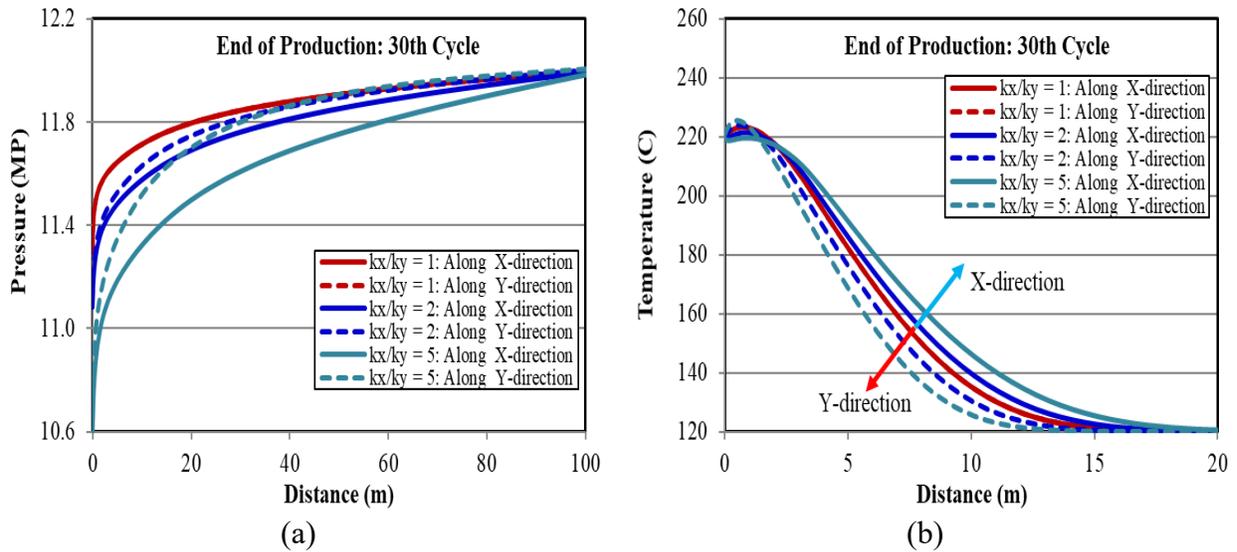

(a)            (b)

**Figure 5**: (a) Pressure and (b) temperature profiles at the end of production in the 30th cycle of operation along the x- and y-axes at mid-height of the formation for different horizontal anisotropies.

The temperature profiles in the x- and y-directions are identical for the isotropic case i.e., $k_x/k_y = 1$ since the absolute permeability is the same in all directions. Similarly, the pressure profiles are also the same in the x– and y- directions in this case. For all of the anisotropic cases, the pressure and temperature fronts in the x-direction move ahead of those in the y-direction. An 11.5 MPa pressure front reaches 20.2 meters in the x-direction and only 9.3 meters in the y-direction for a horizontal anisotropy of $k_x/k_y = 5$. For the same horizontal anisotropy, a 160$^O$C temperature front moves 8.2 meters in the x-direction and only 5.7 meters in the y-direction. Bottom hole pressure and temperature are important for operational purposes. These parameters change with the cyclic operations of the well, as shown in **Figure 6**.

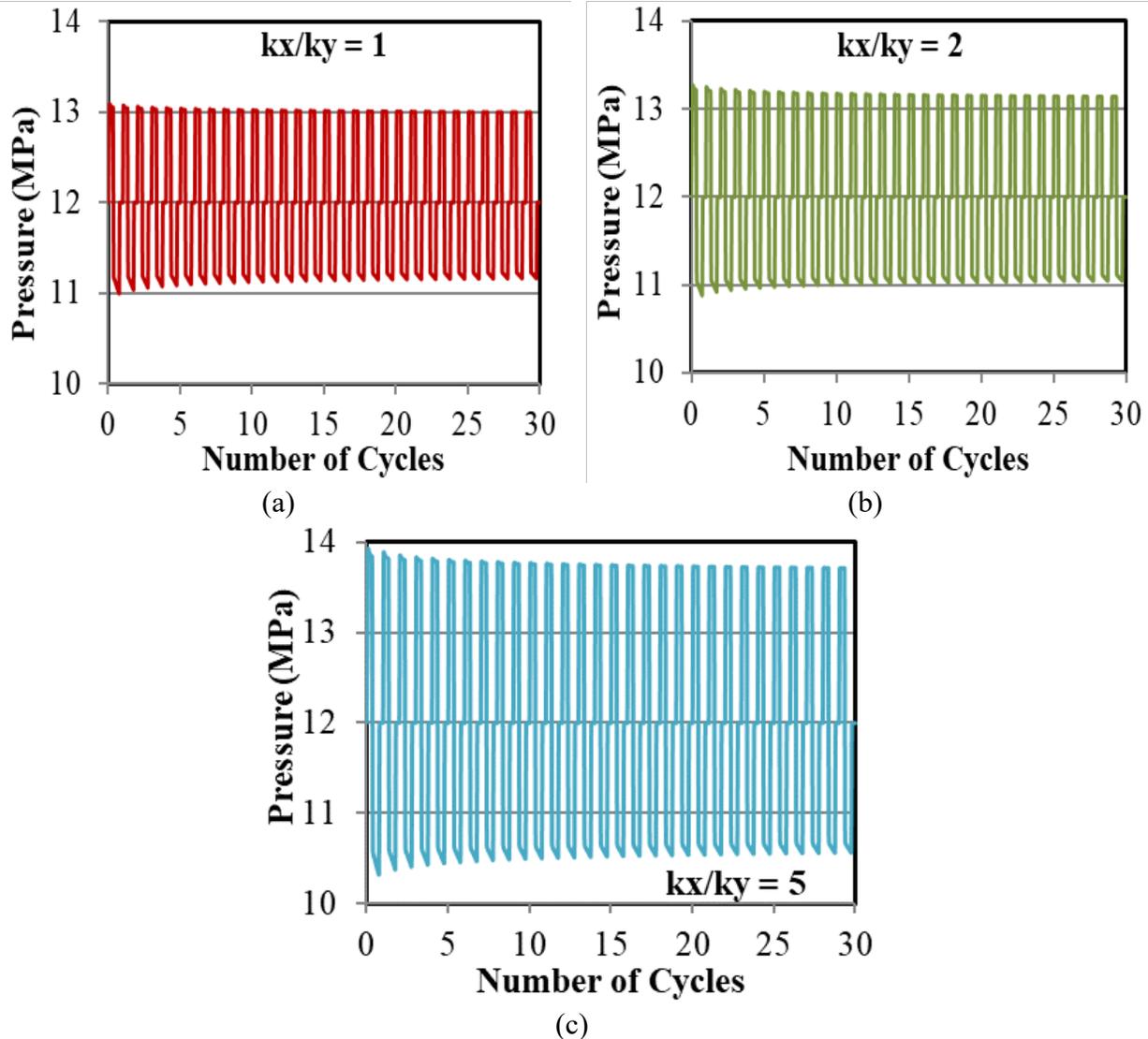

(a)

(b)

(c)

**Figure 6**: Bottom hole pressure at mid-height of the formation versus cycles of operation for horizontal anisotropy, $k_x/k_y$, of (a) 1 (b) 2 (c) 5.

As is well known, the equivalent permeability on the horizontal plane (x-y) is reduced with increasing anisotropy as the permeability in the y-direction is decreased while keeping the permeability in the x-direction fixed at 100 mD. Consequently, higher injection pressure is required for higher anisotropy. Similarly, lower bottom hole production pressure results from the higher drawdown required to produce the same amount of water. The shut-in pressure returns to the initial reservoir pressure of 12 MPa due to the constant pressure boundary condition of 12 MPa.

The temperature profiles do not change significantly with horizontal permeability anisotropies. This profile similarity is because the thermal conduction is independent of permeability. Besides, the thermal equilibrium is a slow process and the daily cycles of 8 hours of injection, 10 hours of production and 6 hours of shut-in are not sufficient for slow heat transfer to alter the water temperature significantly. The bottom hole temperature at the end of injection is always constant at the injection temperature of $250^O$C because the injected water stays near the wellbore. The average production temperature slowly increases with each operational cycle. As more hot water is injected, the formation gets heated with each cycle. Therefore, less amount of heat is lost in the formation due to less temperature gradient between injected water and the rock. It is also noticed that the bottom hole temperature at the end of production is marginally higher for higher horizontal anisotropy in permeability.

**Effect of Layered Heterogeneity**

Five cases are selected by varying the permeability of each layer as shown in **Table 2**.

**Table 2:** The permeabilities of 11 layers of the formation (middle segment) for five cases for layered heterogeneity study

| Vertical Layers of Formation | Permeability, $K_x$ and $K_y$ (mD) | | | |
|---|---|---|---|---|
| | Case 1 | Case2 | Case 3 | Case 4 |
| 1 | 190 | 160 | 130 | 180 |
| 2 | 172 | 148 | 124 | 4 |
| 3 | 154 | 136 | 118 | 180 |
| 4 | 136 | 124 | 112 | 4 |
| 5 | 118 | 112 | 106 | 180 |
| 6 | 100 | 100 | 100 | 4 |
| 7 | 82 | 88 | 94 | 180 |
| 8 | 64 | 76 | 88 | 4 |
| 9 | 46 | 64 | 82 | 180 |
| 10 | 28 | 52 | 76 | 4 |
| 11 | 10 | 40 | 70 | 180 |

Cases 1 to 3 have permeabilities in the descending order from the top layer to the bottom layer. On the other hand, in case 4, alternating layers have high and low permeabilities. Case 1 has the highest gradual variability and case 3 has the lowest gradual variability. To compare results on the same basis, the average permeabilities of all five cases are kept the same at 100 mD.

We have first analyzed the flow distributions of injected and produced water in different horizontal layers to understand the temperature and pressure profiles in a reservoir with layered heterogeneity. Injectivity or productivity of a well particularly a perforation depends mainly on two factors; the permeability of the layer and the drawdown pressure ( i.e., difference between reservoir pressure and bottom hole pressure). Higher bottom hole pressure pushes more water in the formation in case of injection and higher permeability of a layer allows more water to pass through the media with lesser resistance. Descriptions of 4 different cases of layered heterogeneity are provided in Table 2. The homogeneous case is considered as case 5 to compare results. Distributions of mass flow rates of injected water among 11 perforated layers for 5 cases of layered heterogeneity after the 30th cycle of operation are shown in **Figure 7**.

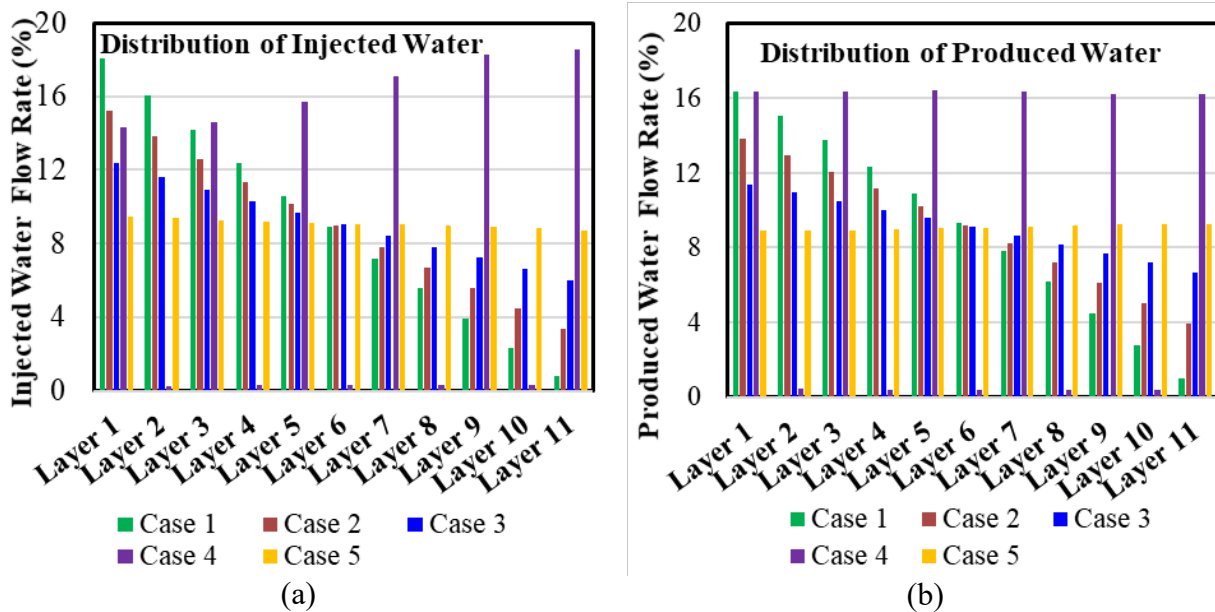

**Figure 7.** Comparisons of distribution of mass flowrates in 11 perforated layers of the formation after the 30th cycle of operation for different layered heterogeneities (a) injected water (b) produced water

In case 1, the top layer has the highest permeability and it gradually decreases towards the bottom layers (Table 2). Therefore, more water (18%) is injected in the top layer and about 1% in the bottom layer. Although the wellbore at the bottom layer has the highest pressure due to added hydrostatic pressure, the effect is mitigated over the effect of permeability of the layer in this case. The same results are observed for cases 2 and 3 also. In case 5 which is a homogeneous case, the injected water is almost equally distributed among the layers as expected. Case 4 of alternating high and low permeability layers has an interesting distribution of injected water. All high permeability layers received 14-19% of water whereas the low permeability layers received only

0.3% each. The bottom layer had the highest amount of water injection (19%) due to the higher bottom hole pressure and high permeability compared to the permeability of the nearest layer. The distributions of hot injected water among different layers have direct implications on the temperature distributions in the reservoir as shown in **Figure 8**.

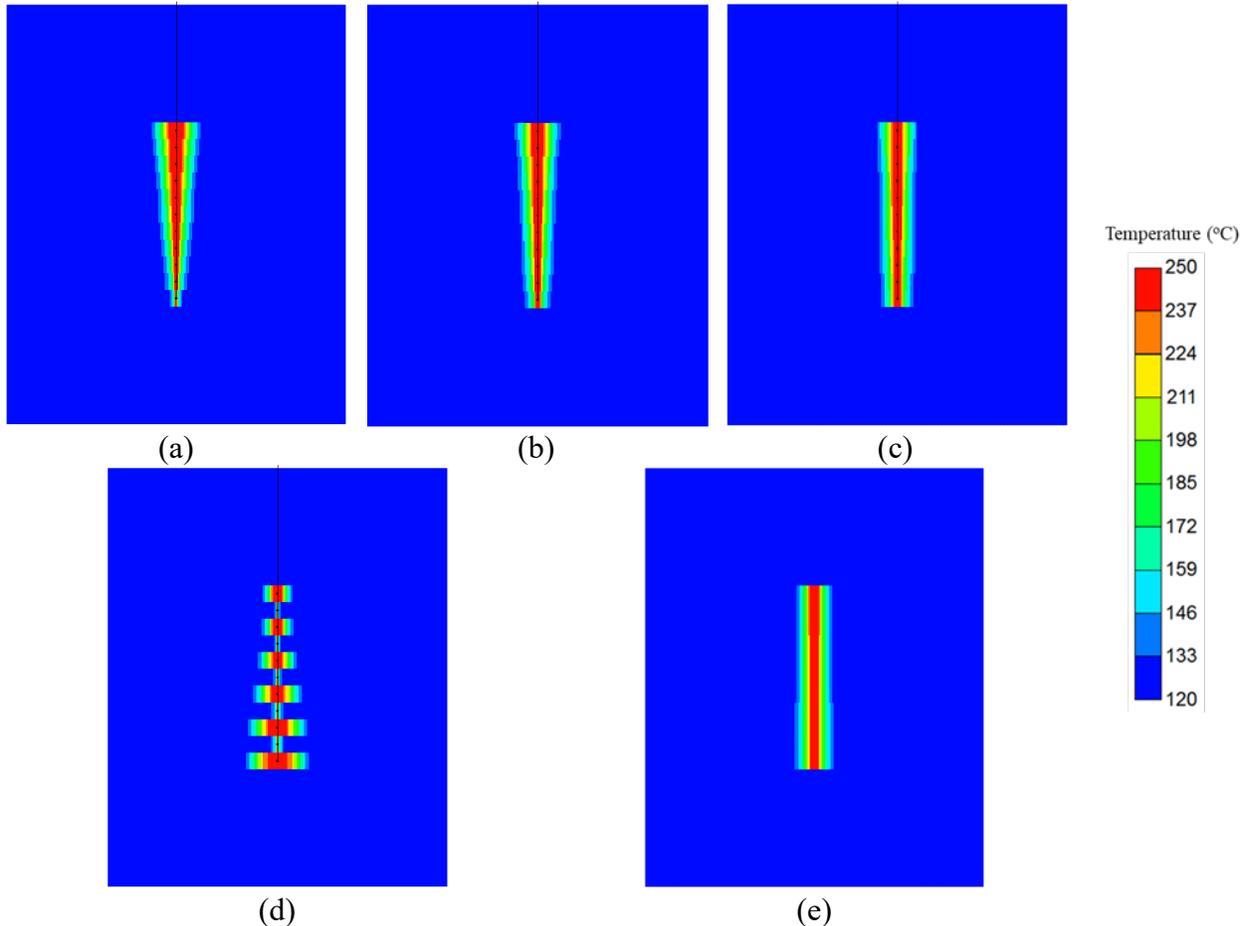

(a) (b) (c)

(d) (e)

**Figure 8**: Spatial distribution of temperature (°C) on a vertical plane (x-z plane crossing well perforations, 200 meters X 250 meters) at the end of injection of the 30$^{th}$ cycle of operation for layered heterogeneities (a) Case 1 (b) Case 2 (c) Case 3 (d) Case 4 (e) Case 5

The temperature profiles of all cases follow the distributions of injected water at 250°C. The wellbores have the same temperature as the temperature of injected water. The temperature of the nearby region of wellbore also remains at around 250°C due to poor mixing of injected water with in-situ water at 120°C. The in-situ water is pushed farther from the wellbore. However, the mixing front is only about a few meters to 30 meters away from the wellbore. Beyond the mixing front, the reservoir temperature remains at the initial temperature of 120°C. It is about 30 meters in the bottom layers in case 4 and it is even less than 1 meter in the bottom layer in case 1 where the permeability of the bottom layer is very low (10 mD).

The distributions of produced water from different layers as shown in figure 7 (b) resonate with the results of the distributions of injected water (Figure 7(a)). Except for case 4, the produced water is proportional to the injected water in each layer. Equal amounts of water are produced from each high permeability layer in case 4. Similar results are obtained from the low permeability layers too ( case 4). This is probably due to the bottom hole pressure in the upper layers are less. Therefore, despite the higher injection ( not very significant though) in the bottom layer, the production of water was not produced at the same rate due to higher bottom hole pressure.

The temperature spreads after the production of the 30$^{th}$ cycle are quite similar to the temperature profile after injection except the temperatures are lower. During production, hot water near the wellbore is mostly produced. Therefore, the temperature drops around the wellbore and the mixing fronts moves towards the wellbore. The injection of hot water at 250°C and the production of the same amount of water from the reservoir thereafter have an impact on the recovery of heat in the geothermal battery system. Cumulative energy recovery versus the number of cycles (or days) for all five cases of layered heterogeneity is shown in **Figure 9**.

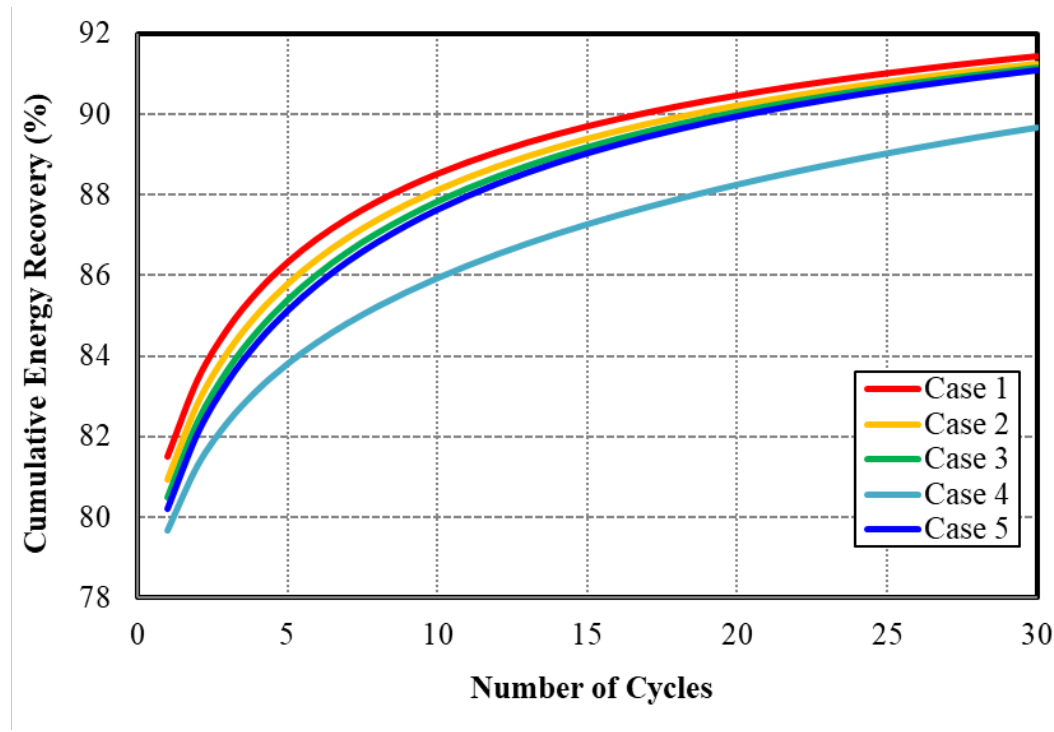

**Figure 9**. Cumulative energy recovery with the number of cycles for different types of layered heterogeneity

The highest amount of heat is recovered (91.4%) in case 1 where the top layer has the highest permeability and the gradual decline of permeability in each lower layer is the maximum. Due to the mixing of injected hot water (250°C) with relatively cold in-situ water (120°C) heat is always lost even if the same amount is produced as injected which is happening for cases 1 to 3 and 5. As described earlier that the maximum of hot water is injected (18.0%) and produced (16.2%) to/from the top layer. The difference in the injection and production is even less for other layers. Therefore, the loss of hot water (i.e., a higher amount of heat) is less. The loss increases as the difference in permeabilities among layers diminishes. However, the differences in cumulative heat recoveries after the 30$^{th}$ cycle for cases 1 (91.4%), 2(91.3%), 3(91.2%) and 4(91.1%) are not significant. In case 4, the cumulative heat recovery after the 30$^{th}$ cycle is 89.7%. This is because the 18.5 % hot water (250°C) is injected in the bottom layer (layer 11) and 16.2 % mixed water having a temperature between 120°C and 250°C is produced. In the best-case scenario, around 2% of in-situ cold water (120°C) is mixed and produced. On the other hand, 14.3 % of hot water is injected in the top layer (layer 1) and 16.4 % is produced. In both cases, the heat is lost in the reservoir.

**Conclusions**

Horizontal permeability anisotropy has a significant influence on the bottomhole pressure and temperature and their spatial distributions. During injection, the pressure front travels more in the (higher permeability) x-direction compared to the y-direction, progressively so as horizontal anisotropy increases. As is familiar in secondary recovery, this leads to an elliptical shape of the pressure contours, whereas the shape becomes circular for the isotropic case. The thermal front moves slowly to a few meters (around 25 meters only) and the pressure front reaches the boundary (100 meters). Again, this is well known from thermal operations in the oil field.

The pressure reaches an equilibrium value of 12 MPa (initial formation pressure) when the well is shut-in for all anisotropic cases (this re-equilibration occurs fairly rapidly because of the relatively high permeabilities). However, the temperature remains essentially unchanged during shut-in due to the slow conductive heat transfer.

In case of layered heterogeneity in permeability, distributions of injected and produced water are proportional to the permeability of the layer except for the case of alternating high and low permeability layers. More water is injected in the bottom layer with high permeability compared to the top layer with same permeability. Layered heterogeneity has less effect on cumulative heat recovery in case of gradual decline of permeability from top layer to bottom layer. However, the heat recovery is less (89%) for alternating high and low permeability layers.

**Acknowledgement**

Funding for this project was provided by the U.S. National Science Foundation, under grant NSF-EAGER- 1912670.

**Appendix: Effect of Vertical Anisotropy**

The effects of vertical anisotropy in the permeability; i.e., the ratio of the permeability in the x-direction and permeability in the z-direction ($k_x/k_z$) are summarized. No significant changes in temperature and pressure profiles on any vertical planes (x-z) are observed. Injection, production and shut-in data are shown.

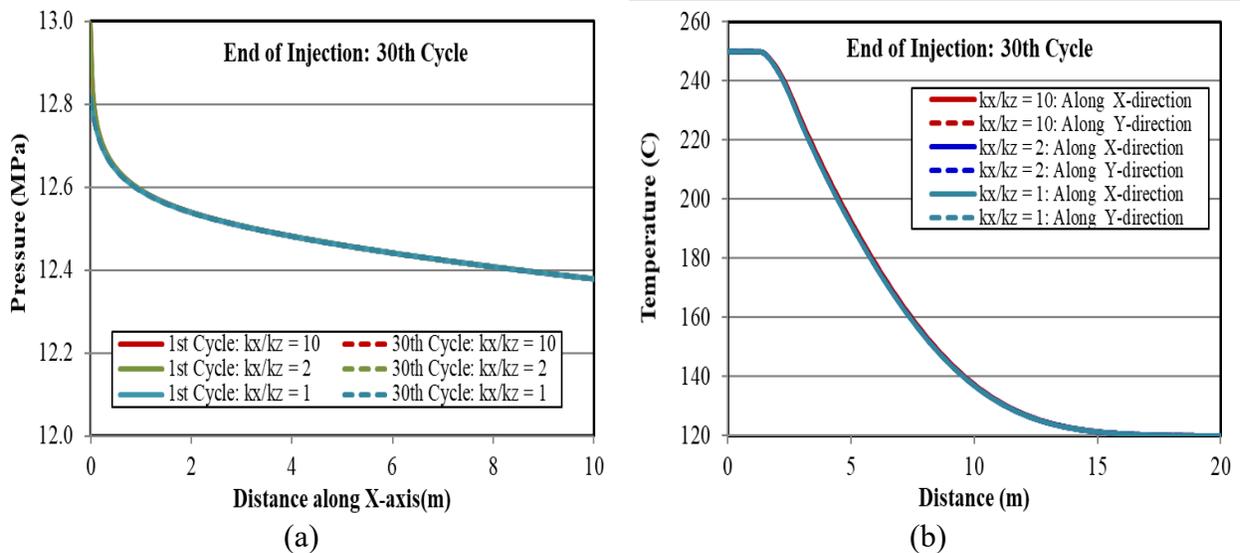

**Figure A.1**: (a) Pressure and (b) temperature profiles at the end of injection FOR the 30$^{th}$ cycle of operation along the x and y axes at the mid-height of the formation for different vertical anisotropies.

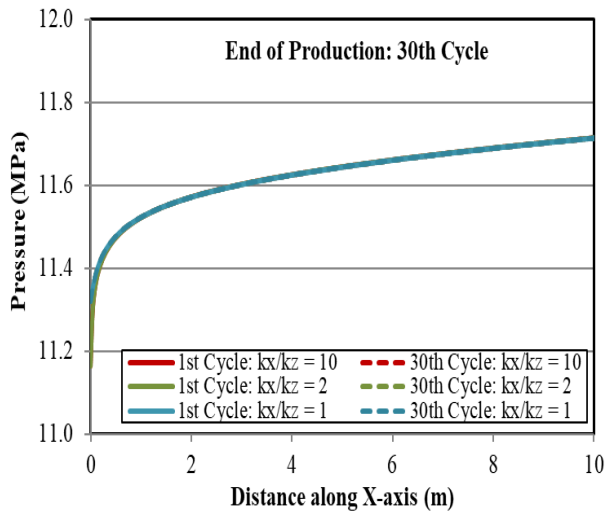 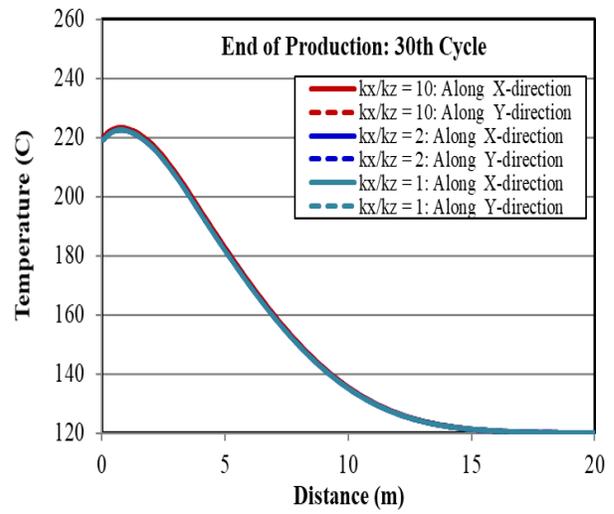

**Figure A.2**: (a) Pressure and (b) temperature profiles at the end of production of the 30th cycle of operation along two perpendicular directions of the x and y-axes at the mid-height of the formation for different vertical permeability anisotropies.

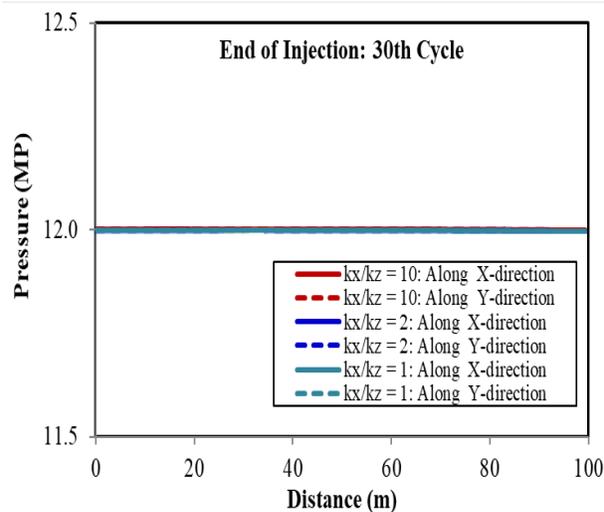 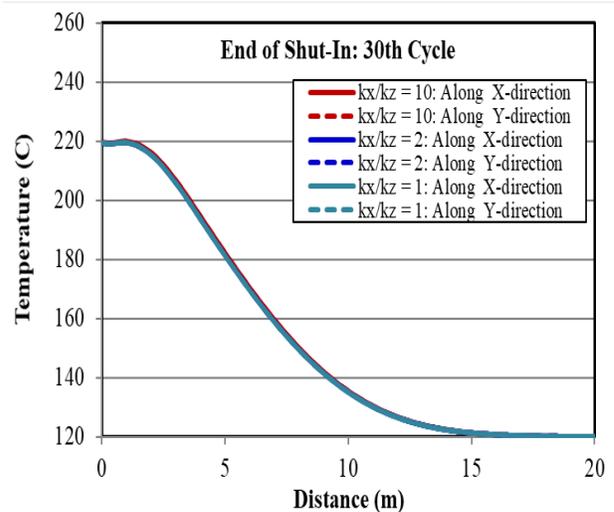

**Figure A.3**: (a) Pressure and (b) temperature profiles at the end of shut-in during the 30th cycle of operation along two perpendicular directions of the x- and y-axes at the mid-height of the formation for different vertical anisotropies.

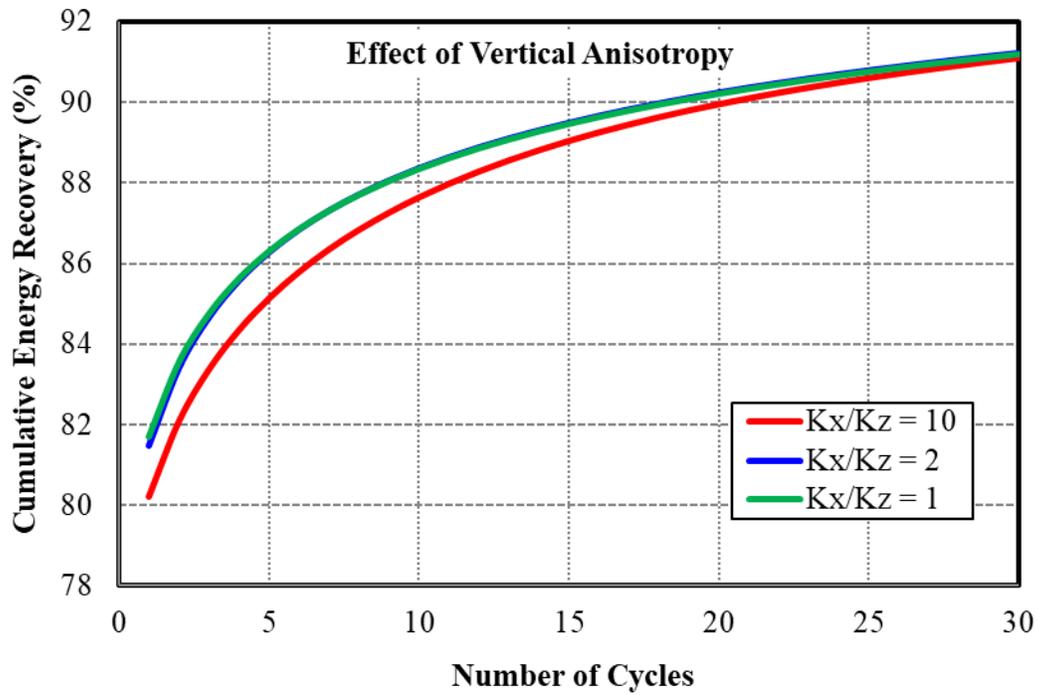

**Figure A.4**: The cumulative energy recovery for three horizontal permeability anisotropy scenarios.

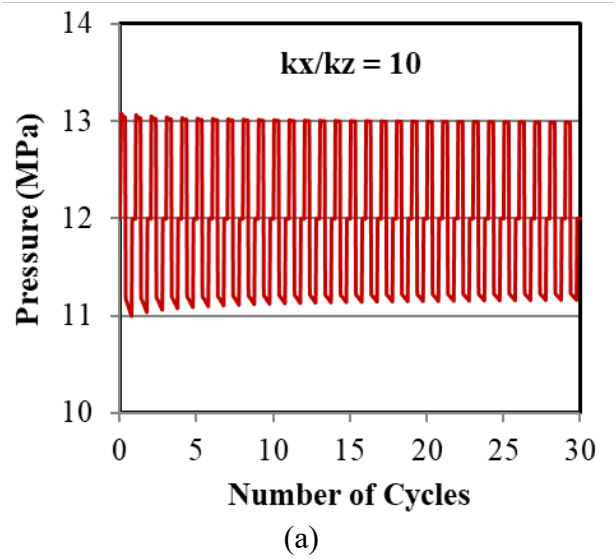
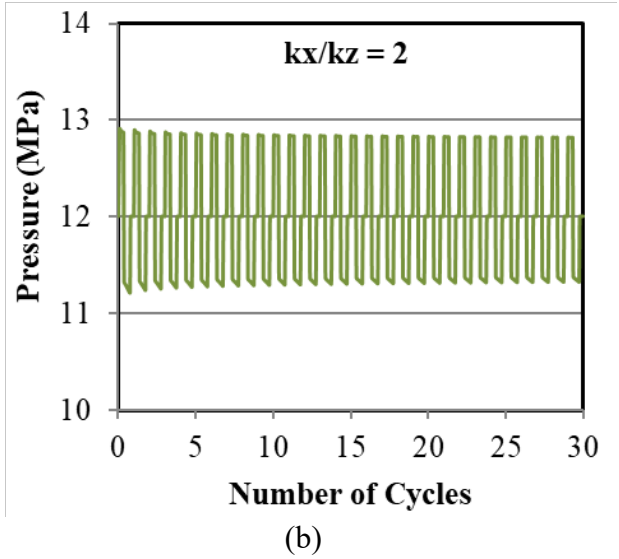

(a)　　　　　　　　　　　　　　　　　　(b)

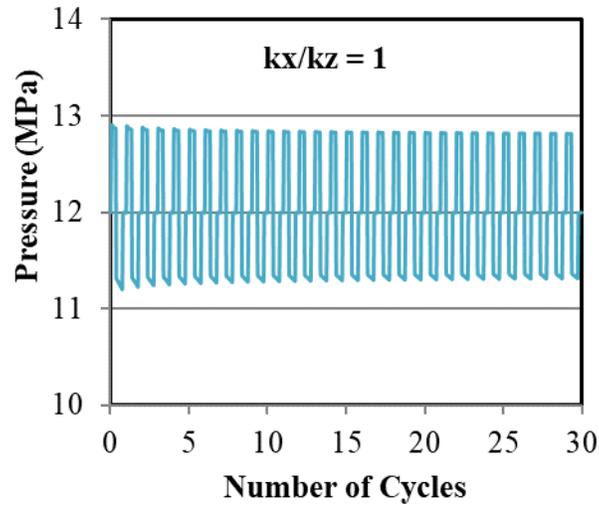

(c)

**Figure A.5**: Bottomhole pressure at the mid-height of the formation versus cycles of operation for vertical anisotropy, $k_x/k_z$ of (a)10 (base case) (b) 2 (c) 1.

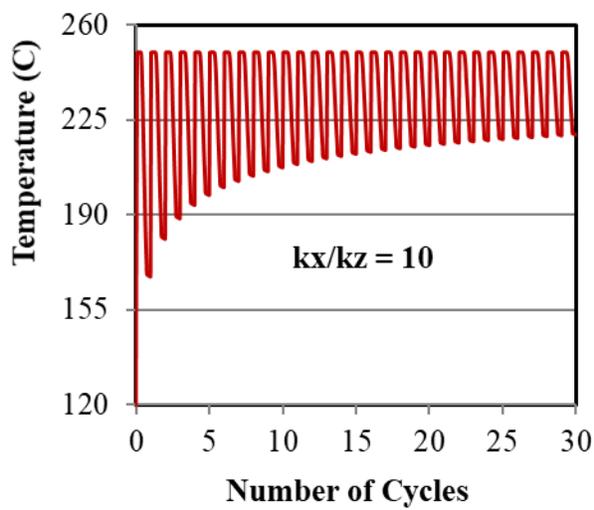

(a)

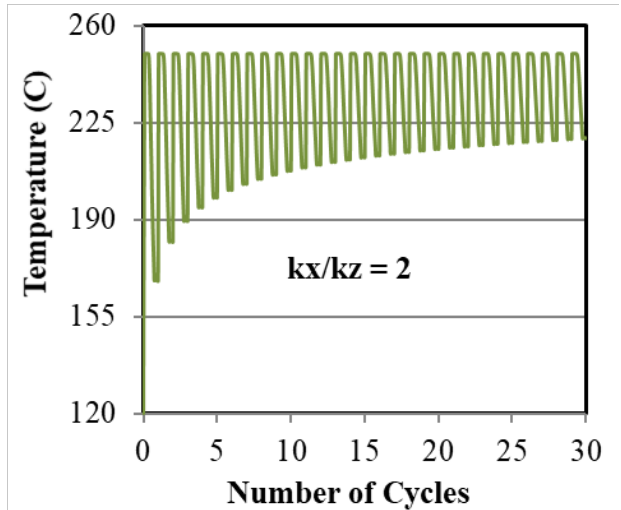

(b)

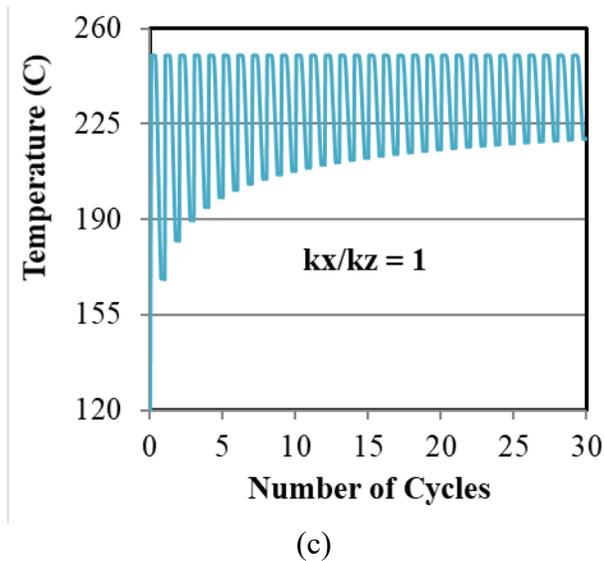

(c)

**Figure A.6**: Bottomhole temperature at the mid-height of the formation versus cycles of operation for vertical anisotropy, $k_x/k_z$ of (a)10 (base case) (b) 2 (c) 1.